**Topological Phase Transitions in a Hybridized Three-Dimensional Topological Insulator**

Su Kong Chong[1], Lizhe Liu[2], Taylor D. Sparks[2], Feng Liu[2] and Vikram V. Deshpande[1]*

[1]Department of Physics and Astronomy, University of Utah, Salt Lake City, Utah 84112 USA

[2]Department of Materials Science and Engineering, University of Utah, Salt Lake City, Utah 84112 USA

*Corresponding author: vdesh@physics.utah.edu

As the thickness of a three-dimensional (3D) topological insulator (TI) becomes comparable to the penetration depth of the surface states, quantum tunneling between surfaces turns their gapless Dirac electronic structure into a gapped surface state. Analytical formulation suggests that the hybridization gap scales exponentially with decrease in number of layers while the system oscillates between topologically trivial and non-trivial insulators. This work explores the transport properties of a 3D TI in the inter-surface hybridization regime. By experimentally probing the hybridization gap as a function of BiSbTeSe$_2$ thickness using three different methods, we map the crossover from the 3D to 2D state. In the 2D topological state, we observe a finite longitudinal conductance at ~$2e^2/h$ when the Fermi level is aligned within the surface gap, indicating a quantum spin Hall (QSH) state. Additionally, we study the response of trivial and non-trivial hybridization gapped states modulated by external out-of-plane magnetic and electric fields. Our revelations of surface gap-closing and/or reopening features are strongly indicative of topological phase transitions (TPTs) in the hybridization gap regime, realizing magnetic/electric field switching between band insulating and QSH states with immense potential for practical applications.

Despite the well-studied topological surface states in 3D TIs [1], their topological 2D limit with quantum spin Hall (QSH) edge states, predicted to exist in ultrathin 3D TIs [2,4], has yet to be realized. Such 2D TIs host a surface gap with inverted bands initiated *via* hybridization of the surface states of the parent 3D TI. From theoretical perspective, the gapped surface spectrum exhibits an oscillatory behavior alternating between topologically trivial and non-trivial 2D states as a function of the layer thickness [2,4]. Although the hybridization gap has been systematically probed in prototypical 3D TIs $Bi_2Se_3$ by angle-resolved photoemission spectroscopy (ARPES) [5,6], and $Sb_2Te_3$ by scanning tunneling microscopy [7], the identified 2D TI phases and their QSH states await to be confirmed electronically in transport measurements. Previous transport measurements in the hybridization regime were restricted by disorder and inhomogeneity in TI thin films preventing the observation of the surface hybridization gap [8,9]. Also, crystal quality of the TI films limited the mobility which constrained the mean free path to microscopic length scales [10]. From the technical aspect, electrical measurements can resolve the prevailing small energy gap scales beyond the resolution limit of ARPES [5,6].

Band inversion through a gap-closing and reopening mechanism is a hallmark of a topological phase transition (TPT) [1]. In addition to the layer-dependent topological phases switching with thickness of 3D TIs, a TPT can be extrinsically induced through band distortions by in-plane or out-of-plane magnetic field [11,12], strain or pressure [13,14], and electric field [7,15-18]. The external in-plane magnetic field can induce gap-closing by oppositely shifting the two surface bands [13]. However, a strong magnetic field is required to fully close the surface gap which is impractical for most applications. Another mechanism with which to transform the band topology is by distorting the crystal lattices, such as strain or pressure induced TPTs in $ZrTe_5$ [13] and BiTeI [14], respectively. Finally the most preferred route is a TPT driven by perpendicular electric fields, which can realize a functional all-electrical topological switch between normal and inverted gap states by controlling the electric potential between top and bottom surfaces [16,17].

**Topological phase diagram.** We first perform the first-principles calculations on a Bi$_{0.7}$Sb$_{1.3}$Te$_{1.05}$Se$_{1.95}$ (BSTS) 3D TI *via* a density functional theory (DFT). The hybridization between the top and bottom surface states of the BSTS causes a tunneling gap opening at and below 10 quintuple layers (QLs) (Fig. S1). The parity of the hybridization gap is evaluated in Table S1. The negative parity of the gap occurs when the hole sub-band is at higher energy level than the electron sub-band, and thus the total parity reads negative, analogous to the description of the negative gap due to bulk band inversion in 3D TIs. Similar to the binary Bi$_2$Se$_3$ and Bi$_2$Te$_3$ 3D TI compounds [2-4], the parity of BSTS hybridization gap exhibits an oscillatory pattern, indicating a switching between band and QSH insulators via thickness modulation as shown in Fig. 1a.

The inter-surface hybridized BSTS exhibits unique electromagnetic responses. In a perpendicular magnetic field, the surface states' hybridization essentially lifts the degeneracy of the N= 0 surface Landau level (LL), and causes a splitting into electron and hole-like N= 0 sublevels [19]. The hybridization gap can thus be defined as the energy difference between the two sublevels, denoted by $E_0$. The analytical formula of $E_0$ reveals a linear relation with magnetic field, expressed as: $E_0 = \tilde{C}_0 + \frac{eB}{\hbar}\tilde{C}_2 - \frac{\mu_B g B}{2}$ [20], where $\mu_B$ and g are the Bohr magneton and effective g-factor; $\tilde{C}_0$ and $\tilde{C}_2$ are the parameters related to the surface states' Hamiltonian. Meanwhile in the presence of external perpendicular electric field, the surface states' Hamiltonian is again modified by the electric field as $H_U = e\mathbf{E}_z \cdot \mathbf{z}$ [15]. The external electric field can induce a Rashba-type splitting in surface band [15,16] due to the structure inversion asymmetry in the gapped surface states, resulting in a shrinkage of the hybridization gap.

Combining the responses of the hybridization gap to magnetic and electric fields, we construct phase diagrams for hybridized BSTS with trivial and non-trivial topological phases, based on analytical formulations, as illustrated in Fig. 1b. The figures investigate and compare the switching of topological phases in two thicknesses with distinct topology. For BSTS with inverted surface gap, both perpendicular magnetic and electric fields are accountable for termination of the QSH gap by breaking the time-reversal and inversion symmetries, respectively. The critical switching fields, denoted as $B_c$ and $E_c$, draw the maximum magnetic and electric fields for the QSH state at which boundaries the surface gap vanishes. In contrast, the normal surface hybridization gap reveals noteworthy phase switching only in electric field, while the out-of-plane magnetic field results in monotonic widening of the surface gap [20]. The closure of trivial gap and reopening of inverted gap as controlled by electric field is another route to a QSH state (Fig. S3).

**Hybridization gap.** Experimentally we have employed three different methods to evaluate energy gaps in the hybridized BSTS. The first method is thermal activation energy by fitting the temperature dependent conductance. Fig. 2a shows the $R_{xx}$ as a function of $V_{bg}$ measured at different temperatures for a 9 nm BSTS. The amplification of $R_{xx}$ peak with the decrease in temperature implies an insulating state developed at charge neutrality point (CNP) due to the inter-surface hybridization. The inset of Fig. 2a displays the $G_{xx}$ data taken at the CNP on an Arrhenius plot, where strongly thermally activated conductance is observed. Linear fitting to the activation relation $G_{xx} = G_{xx}^0 \exp(E_A/kT)$ [21] yields an activation energy ($E_A$) of ~6.5 meV. The second method is to probe the non-linear current-voltage characteristics by measuring the differential conductance (dI/dV). Fig. 2b (inset) displays a dI/dV map as functions of $V_b$ and $V_{bg}$ for the 9 nm BSTS. As the $V_{bg}$ sweeps across the CNP, the dI/dV reaches an overall minimum, which elucidates the nature of the surface gap. The diamond-shaped feature arises from charge transport across the surface states when the chemical potential aligned to or detuned from the hybridization gap at the CNP [22]. A hybridization gap of ~18 meV is determined from the dI/dV *versus* $V_b$ plot (Fig. 2b) across the center minimum. The third approach is to determine the chemical potential relation, μ(n), by integrating the reciprocal quantum capacitance, $1/C_Q$, with respect to charge density (n) [23]. The capacitance dip at the CNP (n≈ 0 /cm$^2$) (Fig. 2c) indicates a minimum density of state (DoS) corresponding to the insulating state. The μ(n) plotted in Fig. 2c reveals a step feature, where the step height of ~21 meV gives the hybridization gap for the 9 nm BSTS.

The hybridization gap size as a function of BSTS flake thickness extracted from the three methods are summarized in Fig. 2d. The details of gap extractions for the different thicknesses BSTS are depicted in Fig. S4-S6. Our data shows a nearly exponential dependence of the hybridization gap with the thickness of BSTS, which is in consistent with the results from ARPES [5]. The hybridization gap can be approximated by the relations as $\Delta_h \propto e^{-\lambda d}$ [24], where the characteristic length (λ) is fitted to ~0.6 nm$^{-1}$. We note that our observed single particle tunneling gap of BSTS opens at thickness of 10 nm, indicating a higher 2D crossover point than the ARPES analysis for Bi$_2$Se$_3$ (below 6 nm). Similar crossover thickness has recently been reported for the same BSTS compounds [11]. To verify the single particle gap nature, we have included the DFT calculations (Fig. S1) based on molecular structure of BSTS as depicted in Fig. 2d. The calculated hybridization gap sizes at the thickness range are in agreement with our experimental values. The thermal activation gap in general shows a smaller value compare to the gap size deduced from the differential conductance and electronic compressibility because of its sensitivity to smearing effect by disorders [25]. This wider thickness range of the gapped state provides more accessible number of quintuple layers of 3D TI to probe the possible TPTs and QSH state in the surface gap.

**Normal and inverted gaps.** From the DFT analyses on parity of the hybridization gap (Table S1), we have noted two regimes of inverted surface gap at thicknesses of 9 QLs (<10 meV) and 5-6 QLs (30-40 meV). The transport properties in the first inverted regime (9 and 10 nm BSTS) are studied. Different from the monotonically increasing $R_{xx}$ for the 9 nm BSTS, the 10 nm one shows $R_{xx}$ maximum saturating at ~12 kΩ (~$h/2e^2$) at temperature below 50 K. This weak temperature response in the 10 nm BSTS at low temperature suggests a finite conductance state existing/developing in the hybridization gap [26]. The full data set of $R_{xx}$ at all temperatures for the 9 nm and 10 nm hybridized BSTS as presented in color maps (inset of Fig. 3a) further elaborate their distinct temperature dependence behaviors. Fig. 3b compares the gate-dependence of $R_{xx}$ at temperature of 1.5 K for the 9 nm and 10 nm BSTS. The $R_{xx}$ reaches a maximum value of ~500 kΩ (~$h/0.05e^2$) as the chemical potential is tuned into the surface gap, indicating a normal insulating state for the 9 nm BSTS. Contrary to the highly resistive signal, the 10 nm BSTS exhibits a finite resistance of ~$h/2e^2$ within the hybridization gap regime as indicated by its dI/dV color map inserted in the figure. This again supports the inverted band of the hybridization gap for the 10 nm BSTS where the gap is bridged by a linear-dispersive edge state which can host edge mode of QSH effect [27].

**Magnetic field response.** The normal and inverted surface gaps can be resolved by analyzing their behaviors in perpendicular magnetic field. Line profiles in Fig. 3c plot the $\sigma_{xx}$ and $\sigma_{xy}$ as a function of $V_{bg}$ measured at a magnetic field of 18 T for the 9 nm BSTS. The two N= 0 LL bands developed in the $\sigma_{xx}$ plot are assigned to the electron and hole band edges of the hybridization gap as illustrated by the schematic of the surface band diagram in the inset of Fig. 3c. This splitting of N= 0 LLs is a key signature of the Landau quantization of inter-surface hybridization [19,20]. The established $v$= 0 plateau in the $\sigma_{xy}$ plot within the N= 0 LL, together with the development of $v$= -1 and +1 QH plateaus symmetrically about zeroth plateau, further supports the LLs of the hybridized surface states. Color map of $\sigma_{xx}$ in Fig. 3c illustrates the development of the N= 0 LLs as a function of magnetic field. Similar behaviors were also observed for the 8 nm BSTS (Fig. S7). The N= 0 LL energy spacing ($E_0$) derived from the μ(n) relation shows a linear magnetic field dependence (Fig. S8), in agreement with the analytical model [20]. The linear fitting of $E_0$ to magnetic field yields $\widetilde{C}_2$ of ~70 and ~90 eV Å$^2$ for the 9 and 8 nm BSTS, respectively, which is comparable to the fitting parameters from the surface states' Hamiltonian [28].

The gate-dependent $\sigma_{xx}$ and $\sigma_{xy}$ plots for the 10 nm BSTS measured at the magnetic field of 18 T (Fig. 3d) reveal a similar type of N= 0 LLs splitting. The relatively narrow zeroth LL plateau width in charge density compared to the 9 nm BSTS is consistent with its smaller size of hybridization gap as previously discussed in the gap analyses. Interestingly, the development of the N= 0 LLs with magnetic field traced by the dashed lines along the $\sigma_{xx}$ minimum in Fig. 3d reveal an intriguing feature. The two N= 0 sublevels develop oppositely as revealed by the $\sigma_{xx}$ color map at low magnetic field, signifying the inversion of electron and hole sublevels at zero field. As the magnetic field increases, these two sublevels eventually cross and continue developing into the normal QH states with the two sublevels interchanged. This is equivalent to a closing and reopening of the surface gap to transform from its topologically non-trivial into a trivial gap state. This magnetic field mediated TPT is similar to the band crossing for the QSH effect observed in HgTe/CdTe quantum well [27], and has more specifically been discussed in the theoretical models for ultrathin 3D TIs [12]. In addition, the sublevels crossing feature formed symmetrically about the opposite magnetic field further confirms this magnetic field driven topological switching. The levels crossing point at critical magnetic field ($B_c$) of ~10 T is equivalent to a Zeeman energy of $E_B = g\mu_B B_c$ ~10 meV for g factor of about 20 [29].

**Electric field response.** According to our DFT calculations, an external electric field can also induce TPT by modulating the surface gap. To trigger this effect, we applied a transverse electric field by dual-gating the hybridized BSTS via a vdW heterostructure using two Gr/hBN gate configurations as illustrated in Fig. 4a. The $R_{xx}$ color maps and their line traces in dual-gate voltages for the different thickness ultrathin BSTS are presented in Fig. S9. We note that the CNPs of the hybridized BSTS devices locate at nearly zero $V_{tg}$, inferring that a minimum electric field acts initially on the samples. The diagonal $R_{xx}$ maxima appearing in dual-gating maps justifies the strong coupling between their top and bottom surface states. The diagonal $R_{xx}$ feature, as a result of the opposite polarity in top and bottom gate voltages, behaves very differently in BSTS with variable thicknesses. This lets us infer the dependence of the hybridization gap on the perpendicular electric field.

To better elucidate the electric field response, we converted the dual-gate voltages into displacement field (D) *versus* total charge density (n) relations [30]. The $R_{xx}$ color maps in D and n for the ultrathin BSTS with thickness decreasing from 10 nm to 7 nm are depicted in Fig. 4b. The dI/dV maps of the respective BSTS devices are inserted in the figure to compare the D-dependent features to their hybridization gaps. As discussed in dual-gated transport, the $R_{xx}$ for the 10 nm BSTS responds weakly to the displacement field. Whereas for the 9 nm BSTS, the $R_{xx}$ along the zero-charge density decreases greatly even for small D (several tens of mV/nm). This effect is not affected by the polarity of D as the $R_{xx}$ change is highly symmetric for opposite values of D. However, the D responses of $R_{xx}$ becomes less pronounced as the hybridization gap size increases. To study the D responses of the hybridization gap, we plot the $R_{xx}$ maxima as a function of D taken at n= 0 for the different thickness BSTS in Fig. 4c. As shown in the figure, the $R_{xx}$ falls substantially with D and tends to saturate at large D (> 150 mV/nm) to a value close to the order of $h/e^2$, suggesting a closing of hybridization gap by applying large D. This gap-closing effect is less intense as indicated by the smaller change in $R_{xx}$ with D for the thinner BSTS due to their larger hybridization gap.

To investigate the gap-closing feature by the applied D, we study the temperature response of conductance at different D for the 9 nm BSTS. The activation behaviors are analyzed by plotting $G_{xx}$ *versus* $T^{-1}$ for different D as depicted in Fig. 4d. The insignificant change in $G_{xx}$ with temperature at large D (>150 mV/nm) indicates a negligible thermal activation, thus supports the decreasing trend in D-dependent hybridization gap. As the thermally-activated gap can be affected by smearing effect, we further analyze the µ(n) relations extracted from the dual-gated $C_Q$ map (Fig. S10a) for the 9 nm BSTS. The great reduction in $C_Q$ dip intensity further supports the suppression of the hybridization gap at large D. Likewise, the hybridization gap sizes at different D are assessed quantitatively from the step heights of the µ(n) plots (Fig. 4e) as summarized in top inset of the figure. The $\Delta_h$ versus D plot clearly shows the hybridization gap-closing behavior happens symmetrically about the opposite D. The results are also consistent with the analytical data derived in literature [15,16]. Similar analyses are implemented for 8 and 6 nm BSTS as included in Fig. S10. A hybridization gap-closing strength of ~0.013 e/Å is estimated by fitting the $\Delta_h$–D curve to a linear function. The $\Delta_h$/D as a function of BSTS thickness in the hybridization regime plotted in Fig. 4e (bottom inset) indicates a linear proportionality of the applied electric field with the gap, which agrees well with our calculated value (0.015 e/Å) from DFT (Fig. S3).

The gap-closing in the hybridized surface states provides a strong indication of the perpendicular electric field-induced TPT. As the role of the external electric field is to invert the valence and conduction band states regardless of the initial band topology, the phase transition can occur in both ways, either from trivial to non-trivial topological state or *vice versa* [15-18]. In our 9 nm BSTS, we anticipate a transition from the normal insulating into a 2D topological state through the gap-closing mechanism mediated by the perpendicular electric field. This is supported by the saturating finite resistance observed at large D for the 9 nm BSTS. Whereas for the 10 nm BSTS, the gap closing feature at large electric field can be inferred from the gradual loss of its quantization as $R_{xx}$ deviates from $h/2e^2$ at large D. However, for both cases, the gap-reopening feature is less pronounced in our measurements, presumably due to the slower gap-reopening rate as indicated by our DFT calculations and literature [15].

In summary, the finite-size effect of ultrathin BSTS 3D TI in the quantum tunneling regime was studied. We reported an exponential decay in the hybridization gap size with increment in flake thickness as evaluated by three experimental methods, and supported theoretically by DFT analyses. The trivial and topological phases in the surface gap were identified by their diverging and finite (~$h/2e^2$) resistances, respectively, in response to temperature and gate voltage. By studying the development of two split N= 0 LL sub-bands in perpendicular magnetic field, we observed a crossing of the LL sub-bands for the inverted surface gap BSTS, implying a transition from QSH to normal QH states. In addition, we verified a gap-closing feature, together with a saturating conductance for the normal insulating gap BSTS in perpendicular electric field. These observations serve as compelling signatures for topological phase transitions in the hybridized 3D TI.

**Methods**

**DFT calculations.** Our calculations were performed within the framework of density functional theory as implemented in the Vienna ab initio simulation (VASP) [31]. The projector augmented wave potentials were adopted with the generalized gradient approximations of Perdew-Burke-Ernzerhof exchange-correlation functional [32], and the cutoff energy was set to 520 eV. The relaxation is performed until all forces on the free ions converge to 0.01 eV/Å and the Monkhorst-Pack k-point meshes of 10×10×1 were used, which have been tested to be well convergence. The vacuum space is at least 20 Å, which is large enough to avoid the interaction between periodical images. The different BSTS systems were treated with virtual crystal approximation. The van der Waals interaction is described by DFT-D3 method [33]. In addition, the spin-orbit-coupling was included in the calculations of the electronic structure.

**Device fabrications.** Variable thickness of ultrathin BSTS crystal flakes were exfoliated from the bulk crystal [34,35] and then transferred using a micromanipulator transfer stage into the heterostructures of Gr/hBN sandwiched layers. We fabricated the hybridized BSTS devices into the Hall bar configuration with Cr/Au (2 nm/60 nm) as the contact electrodes. The top and bottom Gr/hBN layers serve as the gate-electrode/dielectric layers for applying of perpendicular electric field to the hybridized BSTS (Fig. 1c). Thicknesses of the BSTS flakes were measured by a Bruker Dimension Icon atomic force microscopy. The device dimensions were obtained from the images taken by a high-resolution optical microscope. BSTS devices with thickness ranging from 10 nm down to 1 nm were fabricated and studied. The device specifications are presented in Table S2 (Supplementary).

**Measurements.** Low temperature electrical transport measurements were performed in a helium-4 variable temperature insert with a base temperature of 1.6 K and magnetic field up to 9 T. Four probe resistances of the devices were measured using a Stanford Research SR830 lock-in amplifiers operated at a frequency of 17.777 Hz. The devices were typically sourced with a constant AC excitation current of 10-20 nA. Two Keithley 2400 source meters were utilized to source DC gate voltages separately to the top and bottom Gr gate electrodes. Variable temperature transport measurements were controlled by a Lakeshore temperature controller. The differential conductance was measured using a Stanford Research SR830 lock-in amplifier coupled to a model 1211 current preamplifier. The devices were sourced with an AC excitation voltage of 100 μV.

The DC bias voltage in a range from 40-1200 mV was swept across the source-drain electrodes. Magnetotransport measurements at high magnetic field is carried out in a helium-3 variable temperature insert at a base temperature of 0.3 Kelvin and magnetic field up to 18 tesla based at the National High Magnetic Field laboratory. Two synchronized Stanford Research SR830 lock-in amplifiers were used to measure the longitudinal and Hall resistances concurrently on the BSTS devices. Capacitance was measured in a capacitance bridge configuration [23] connected between the BSTS device and the parallel gold strip as a reference capacitor. Two synchronized (at a frequency of ~50-70 kHz) and nearly equal-amplitude AC excitation voltages (range of 15-40 mV) were applied separately to the top and bottom Gr gates, whose relative magnitude was chosen to match the ratio of geometric capacitances of top and bottom surfaces. A third AC excitation voltage was applied to the reference capacitor with the amplitude set to null the measured signal. The reference capacitors were calibrated to be ~300-400 fF using a standard capacitor (Johanson Technology R14S, 1 pF). The capacitance data were acquired by monitoring the off-balance current at the balance point as the DC gate voltages were changed.

**Acknowledgements**

This material is based upon work supported by the National Science Foundation the Quantum Leap Big Idea under Grant No. 1936383. A portion of this work was performed at the National High Magnetic Field Laboratory, which is supported by National Science Foundation Cooperative Agreement No. DMR-1644779 and the State of Florida.

**Author contributions**

S.K.C. and V.V.D. designed, conducted the experiments and prepared the manuscript. T.D.S. provided single crystal BiSbTeSe$_2$ three-dimensional topological insulator. L.L. and F.L. performed theoretical calculations to support the experimental data. All authors contributed to the discussion of results and approved the final version of the manuscript.

**Additional information**

**Supplementary Information** accompanies this paper at http://www.nature.com/

**Competing financial interests:** The authors declare no competing financial interests.

# Figures

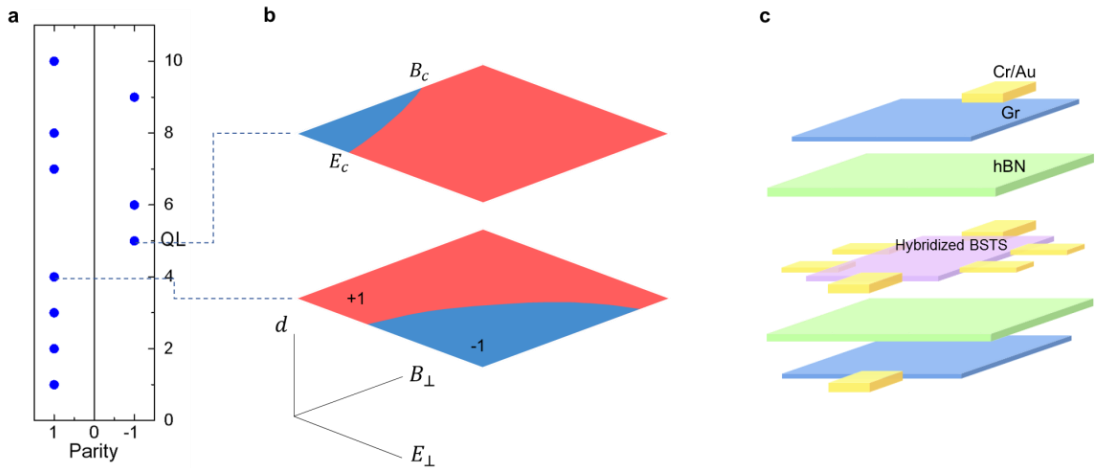

**Figure 1** Topological phase transitions. **a** Plot of the parity of the hybridization gap *versus* thickness of the BSTS 3D TI calculated by DFT. **b** The topological phase diagrams for parity of the surface hybridization gap modulated by external magnetic and electric fields acting perpendicular to the c-axis plane of the ultrathin BSTS. The diagrams are drawn based on the Hamiltonian models described in the references [15,20]. The red and blue colors represent the opposite sign of parity as a representation of the normal and inverted surface hybridization gaps. **c** A schematic of the hybridized BSTS 3D TI device fabricated in a dual-gating configuration for the studies.

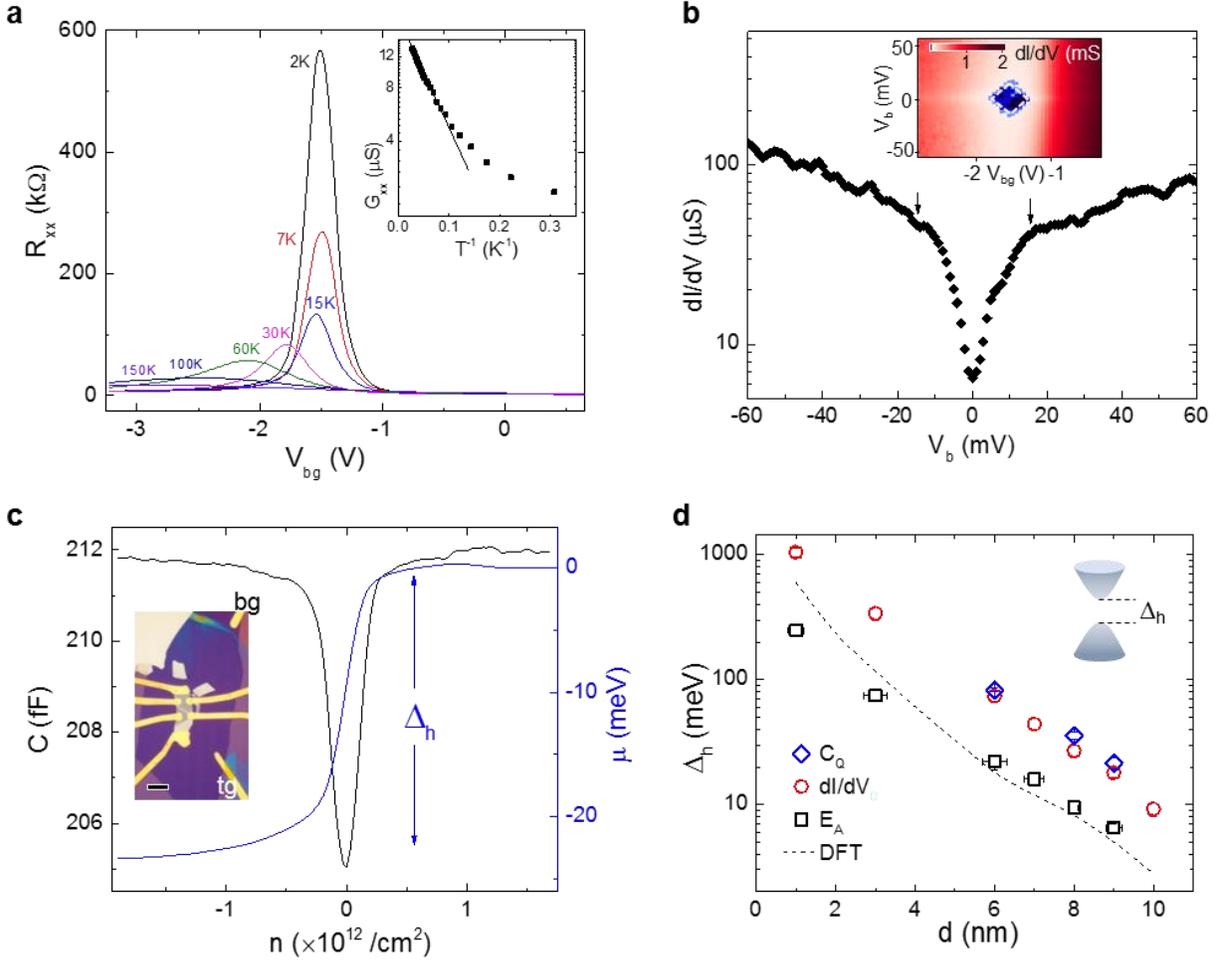

**Figure 2** Probing hybridization gap. **a** $R_{xx}$ *versus* $V_{bg}$ plots at different temperatures. Inset in **a** is an Arrhenius plot of $G_{xx}$ *versus* $T^{-1}$ for $V_{bg}$ fixed at the CNP. **b** dI/dV *versus* $V_b$ curve at insulating region ($V_{bg}$~ -1.5 V). Inset in **b** is a color map of dI/dV as functions of bias voltage ($V_b$) and $V_{bg}$. **c** C and μ(n) as a function of n induced by $V_{bg}$ with $V_{tg}$ fixed at the overall CNP. Inset in **c** is optical image of an ultrathin BSTS vdW heterostructure device. Scale bar in **c** inset is 10 μm. **d** Hybridization gap ($\Delta_h$) obtained from the three different approaches plotted in log scale as a function of BSTS flake thickness. Dashed line in **d** is the $\Delta_h$ extracted from our DFT calculations. Inset in **d** is a schematic of a hybridized surface state.

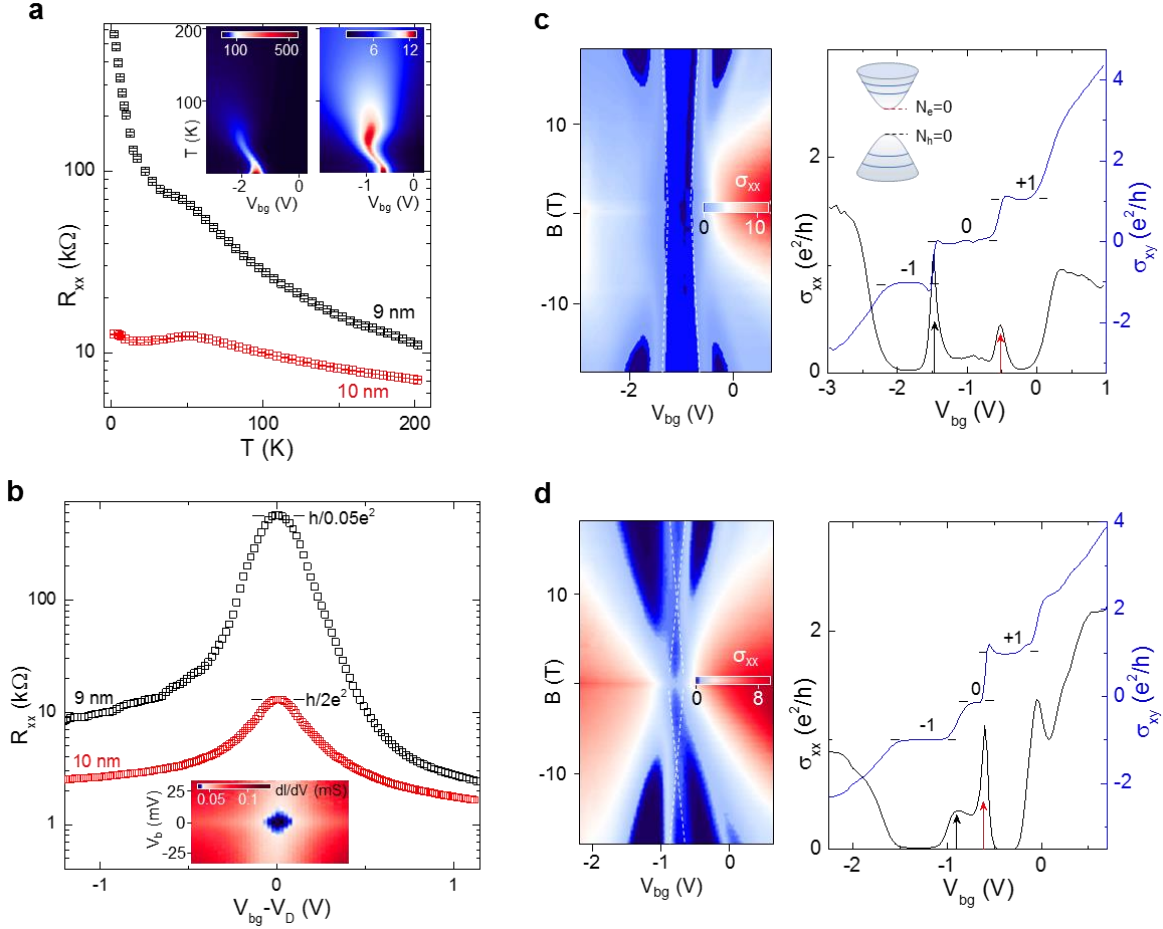

**Figure 3** Normal and inverted gaps. **a** $R_{xx}$ as a function of temperature for the 10 nm and 9 nm BSTS. Inset in **a** are color map of $R_{xx}$ as functions of T and $V_{bg}$ for the 10 nm (right) and 9 nm (left) BSTS. **b** $R_{xx}$ as a function of $V_{bg}-V_D$ for the 10 nm and 9 nm BSTS. Inset in **b** is a color map of dI/dV as function of $V_b$ and $V_{bg}$ for the 10 nm BSTS. Plots $\sigma_{xx}$ and $\sigma_{xy}$ *versus* $V_{bg}$ at the magnetic field of 18 T for the **c** 9 nm and **d** 10 nm BSTS. Inset in **c** is a schematic of the LLs form in the hybridized surface state. Black and red arrows in line profiles in **c** and **d** present the two N= 0 LL bands residing at the hole and electron band edges, respectively. Color maps of the $\sigma_{xx}$ as functions of magnetic field and $V_{bg}$ for the **c** 9 nm and **d** 10 nm BSTS. The white dashed lines in color maps in **c** and **d** trace the development of N= 0 LLs in electron and hole sublevels with magnetic field.

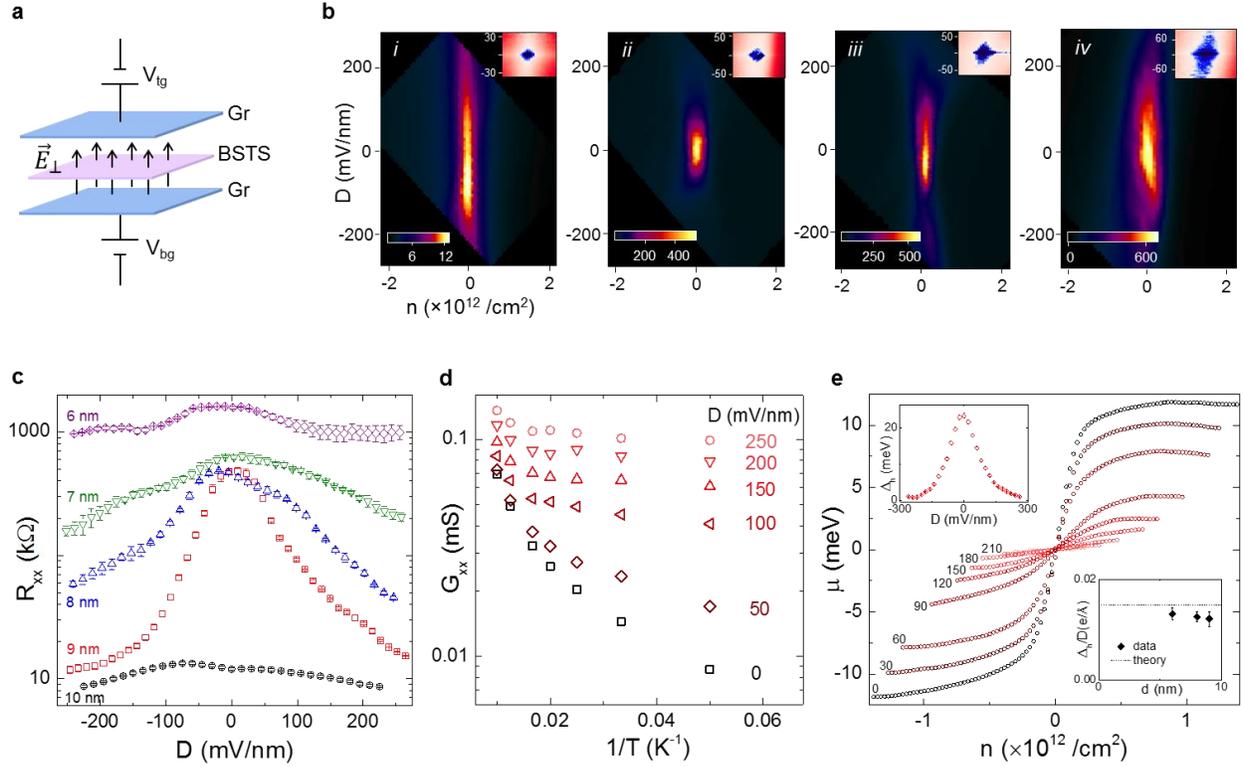

**Figure 4** Gap tuning by external electric field. **a** Schematic of an ultrathin BSTS in a perpendicular electric field induced by the opposite polarity of dual-gate voltages. **b** Color maps of $R_{xx}$ as functions of displacement field (D) and total charge density (n) for the different thickness BSTS in the order of (*i*) 10 nm, (*ii*) 9 nm, (*iii*) 8 nm, and (*iv*) 7 nm. Inset in **b** are the dI/dV color maps as functions of $V_b$ and $V_{bg}$ for the respective samples. **c** $R_{xx}$ as a function of D for the different thickness BSTS. **d** Plots $G_{xx}$ *versus* $T^{-1}$ for the 9 nm BSTS at different D. **e** Plots of μ(n) *versus* n calculated from the $C_Q$ for different D. Inset (top) in **e** is the plot of $\Delta_h$ *versus* D for the 9 nm BSTS. Inset (bottom) in **e** is the plot of $\Delta_h/D$ as a function of BSTS thickness.